\documentclass[conference]{IEEEtran}
\IEEEoverridecommandlockouts
\usepackage{cite}
\usepackage{amsmath,amssymb,amsfonts}
\usepackage{algorithmic}
\usepackage{graphicx}
\usepackage{textcomp}
\usepackage{xcolor}
\usepackage{multirow}

\def\BibTeX{{\rm B\kern-.05em{\sc i\kern-.025em b}\kern-.08em
    T\kern-.1667em\lower.7ex\hbox{E}\kern-.125emX}}
\begin{document}

\title{IRC: Cross-layer design exploration of\\ \underline{I}ntermittent \underline{R}obust \underline{C}omputation units for IoTs}

\author{Arman~Roohi, Ronald F DeMara\\
Electrical and Computer Engineering Department, University of Central Florida,\\ Orlando, FL, 32816-2362, aroohi@knights.ucf.edu

}

\maketitle

\begin{abstract}
Energy-harvesting-powered computing offers intriguing and vast opportunities to dramatically transform the landscape of the Internet of Things (IoT) devices by utilizing ambient sources of energy to achieve battery-free computing. In order to operate within the restricted energy capacity and intermittency profile, it is proposed to innovate Intermittent Robust Computation (IRC) Unit as a new duty-cycle-variable computing approach leveraging the non-volatility inherent in spin-based switching devices. The foundations of IRC will be advanced from the device-level upwards, by extending a Spin Hall Effect Magnetic Tunnel Junction (SHE-MTJ) device. The device will then be used to realize SHE-MTJ Majority/Polymorphic Gate (MG/PG) logic approaches and libraries. Then a Logic-Embedded Flip-Flop (LE-FF) is developed to realize rudimentary Boolean logic functions along with an inherent state-holding capability within a compact footprint. Finally, the NV-Clustering synthesis procedure and corresponding tool module are proposed to instantiate the LE-FF library cells within conventional Register Transfer Language (RTL) specifications. This selectively clusters together logic and NV state-holding functionality, based on energy and area minimization criteria. It also realizes middleware-coherent, intermittent computation without checkpointing, micro-tasking, or software bloat and energy overheads vital to IoT. Simulation results for various benchmark circuits including ISCAS-89 validate functionality and power dissipation, area, and delay benefits. 
\end{abstract}


\section{Introduction}
For the past five decades, complementary metal-oxide-semiconductor (CMOS) has been the dominant technology and it has provided the demanded dimension scaling for implementing high-performance and low-power circuits. The evolution of this charge-based CMOS devices is described and predicted by Moore's law \cite{moore}, this prediction, that number of transistors in integrated circuits (ICs) doubles roughly every two years, has continued for many years. On the other hand, by the inevitable scaling down of the feature size of the CMOS transistors which are deeper in nano-ranges, the CMOS technology has encountered many critical challenges such as high leakage currents, reduced gate control, high power density, increased circuit noise sensitivity and high lithography costs which obstruct the continuous dimension scaling and consequently degrade the suitability of the CMOS technology for the near future high-density and energy efficient applications. Explaining in more details about the first aforementioned restriction, due to the quantum mechanical tunneling of electrons from the gate electrode into the transistor channel through the gate oxide, leakage current occurs. Owing to the Moor's law's and mentioned problems as well as the increasing chip complexity, researchers have to start seeking novel technologies to replace the charge-based devices. Among promising devices, 2015 International Technology Roadmap for Semiconductors (ITRS) \cite{ITRS} identifies nanomagnetic devices as capable post-CMOS candidates. Spintronics devices show promising features such as non-volatility, near-zero static power, and high integration density. The non-volatility means that the data can be maintained even if the power is off, so the standby power is reduced significantly. Moreover, due to the possibility of 3D integration above CMOS designs at the back-end process, distances between logic and memory can be shortened, which reduces considerably the dynamic power. The scalability feature of spin-based devices in addition to their low power characteristic, make the Spintronics as a promising alternative for CMOS architectures.

Recently, magnetic tunnel junction (MTJ) devices are one of the most important component of any spin-based structures, which can be configured into two different stable configurations. Due to the tunnel magnetoresistance (TMR) effect, these two parallel and anti-parallel states show low and high resistance, which can be denoted ``0'' and ``1'' in binary information, respectively. Two of developed switching approaches for MTJs are spin-transfer torque (STT) \cite{armanDW} and spin-Hall effect (SHE) \cite{armanSHE}, in which only one bidirectional current is required. In STT switching approach, the bidirectional current passes through an MTJ according to which it can be configured into P or AP state. Although STT offers several advantages over previous switching methods, it suffers from some challenges such as high write current, and switching asymmetry \cite{SHE-adv2}. Moreover, STT-MTJ is a two-terminal device with a shared write and read path. Consequently, undesirable switching may occur during the read operation, and stored data can be flipped accidentally. Recently, SHE-MTJ, a 3-terminal device, has been researched as a potential alternative offering some benefits such as decoupled read and write paths, as well as energy efficient and high-speed write \cite{SHE-adv3}.
\begin{table*}[h]
\small
\centering
\caption{Selected previous works with significant contributions towards intermittent processor design.}
%\resizebox{\textwidth}{!}{
\begin{tabular}{|c|l|l|c|l|}
\hline
\textbf{Approach} & \multicolumn{1}{c|}{\textbf{Methodology}} & \multicolumn{1}{c|}{\textbf{Features}} & \textbf{Robust Element} & \multicolumn{1}{c|}{\textbf{Challenges}} \\ \hline
\multirow{4}{*}{\begin{tabular}[c]{@{}c@{}}Mementos\\ \cite{mementos}\end{tabular}} & \multirow{4}{*}{checkpointing} & \multirow{4}{*}{\begin{tabular}[c]{@{}l@{}}Run-time energy estimation\\ Periodic system snapshots\\ 340 Byte footprint illustrated\\ V¬ON = 4.5V, V¬OFF = 2V\end{tabular}} & \multirow{4}{*}{Flash} & \multirow{4}{*}{\begin{tabular}[c]{@{}l@{}}New programming \\ paradigms and Languages\\ Data movement overhead\\ Low endurance\end{tabular}} \\
 &  &  &  &  \\
 &  &  &  &  \\
 &  &  &  &  \\ \hline
\multirow{5}{*}{\begin{tabular}[c]{@{}c@{}}Hibernus\\ \cite{duty1}\end{tabular}} & \multirow{5}{*}{\begin{tabular}[c]{@{}l@{}}duty-cycling\\ reactive\\ hibernating\end{tabular}} & \multirow{5}{*}{\begin{tabular}[c]{@{}l@{}}Snapshot before outage\\ 76\%-100\% less processing\\  time and 49\%-79\% lower\\  energy overhead than\\  Mementos\end{tabular}} & \multirow{5}{*}{\begin{tabular}[c]{@{}c@{}}Ferroelectric \\ RAM (FRAM)\end{tabular}} & \multirow{5}{*}{\begin{tabular}[c]{@{}l@{}}NVMs to save all processor \\ register/states\\ Need for sufficient \\ energy to save a full snapshot\\ Long sleeping time\end{tabular}} \\
 &  &  &  &  \\
 &  &  &  &  \\
 &  &  &  &  \\
 &  &  &  &  \\ \hline
\multirow{5}{*}{\begin{tabular}[c]{@{}c@{}}DINO\\ \cite{DINO}\end{tabular}} & \multirow{5}{*}{\begin{tabular}[c]{@{}l@{}}Checkpointing\\ Data versioning\end{tabular}} & \multirow{5}{*}{\begin{tabular}[c]{@{}l@{}}582 Byte footprint \\ illustrated\\ Atomic tasks\\ Reduced flow \\ complexity\end{tabular}} & \multirow{5}{*}{\begin{tabular}[c]{@{}c@{}}Ferroelectric \\ RAM (FRAM)\end{tabular}} & \multirow{5}{*}{\begin{tabular}[c]{@{}l@{}}Large NVM versioning info\\ New programming paradigm\\ Data movement\\ $\sim$4KB average storage \\ overhead\end{tabular}} \\
 &  &  &  &  \\
 &  &  &  &  \\
 &  &  &  &  \\
 &  &  &  &  \\ \hline
\multirow{5}{*}{\begin{tabular}[c]{@{}c@{}}Chain\\ \cite{chain}\end{tabular}} & \multirow{5}{*}{\begin{tabular}[c]{@{}l@{}}task-based \\ control flow\\ Channel-based \\ memory model\end{tabular}} & \multirow{5}{*}{\begin{tabular}[c]{@{}l@{}}Idempotent tasks; \\ no checkpoints\\ 2x to 7.6x performance\\ compared to DINO\end{tabular}} & \multirow{5}{*}{\begin{tabular}[c]{@{}c@{}}Ferroelectric \\ RAM (FRAM)\end{tabular}} & \multirow{5}{*}{\begin{tabular}[c]{@{}l@{}}Hardware redundancy\\ New programming paradigms\\ 42\% larger code than DINO\\ $\sim$8KB average storage \\ overhead\end{tabular}} \\
 &  &  &  &  \\
 &  &  &  &  \\
 &  &  &  &  \\
 &  &  &  &  \\ \hline
\multirow{5}{*}{\begin{tabular}[c]{@{}c@{}}NVP\\ \cite{intermittency1}\end{tabular}} & \multirow{5}{*}{\begin{tabular}[c]{@{}l@{}}PC/register store \\ Partial backup\end{tabular}} & \multirow{5}{*}{\begin{tabular}[c]{@{}l@{}}NV flip-flops in\\  MIPS ISA\\ 1 KHz square waveform\\ 3 MHz clk; 470nF\\  store capacitor\end{tabular}} & \multirow{5}{*}{\begin{tabular}[c]{@{}c@{}}Ferroelectric \\ RAM (FRAM)\end{tabular}} & \multirow{5}{*}{\begin{tabular}[c]{@{}l@{}}Non-volatile internal\\  and external coherence\\ Overhead of checkpointing\end{tabular}} \\
 &  &  &  &  \\
 &  &  &  &  \\
 &  &  &  &  \\
 &  &  &  &  \\ \hline
\end{tabular}
%}
\label{tb1}
\end{table*}

\subsection{Research Motivation}
Intermittent computation approach offers intriguing and vast opportunities to dramatically transform the landscape of IoT devices. The Internet of Thing (IoT) devices require drastically-reduced energy consumption such that they are able to operate using only ambient sources of light, thermal, kinetic \cite{s-all}, and electromagnetic energy as a means to achieve battery-free computing. If lightweight embedded computing could be realized with free and/or inexhaustible sources of energy, new classes of maintenance-free, compact, and inexpensive computing applications would become possible \cite{new-computing}. Thus, energy-harvesting-powered devices could enable a sustainable computing platform for future \cite{medical1}, aerospace \cite{aerospace}, and IoT \cite{IoT} applications. Energy-harvesting devices are projected to develop towards a \$2.6B market by 2024, thus automating human interaction with everyday items in our environment or even life-saving medical implants within our bodies \cite{medical2}.

Therefore, it is proposed to research a promising class of rudimentary processing elements which utilize switching devices capable of leveraging (1) the restricted energy capacity and (2) the intermittent temporal energy profile, of energy harvesting schemes. A typical energy harvesting system converts ambient energy via rectification and charge-trapping. Then energy is accumulated on the capacitor to generate a supply voltage. Once the voltage of the capacitor attains a sufficient level, then a lightweight embedded processing element can commence its operation. However, the stored energy will be rapidly consumed, which consequently precludes the continuation of execution due to an insufficient supply voltage. Hence, the supply voltage of the processor experiences intermittent behavior. This results in an interval, $\tau_{idle}$, of unpredictable unavailability that can interrupt the datapath and the processor clock. This charge/discharge cycle, which is an intrinsic characteristic of energy harvesting devices, may occur more than hundreds of times per second for RF-based sources, and unpredictably for extended durations with kinetic and light-powered sources. Furthermore, the interval $\tau_{idle}$ can occur irregularly and vary in duration leading us to research methods to achieve a new Elastic Model of Computation described herein. Hardware realization of elastic computing addresses one of the major hurdles to the propagation of energy harvesting systems: robust operation despite discontinuities in the ambient energy supplied from its environment.  Robust intermittent operation presents a new and difficult technical challenge that precludes assumptions of conventional processor design of the last several decades. Intermittent behavior can result in disturbances in the execution of programs, data loss, glitch conditions, and lack of execution progress that may lead to irregular and unpredictable results \cite{intermittency1}. Therefore, most of the existing energy harvesting systems are envisioned for rudimentary signal detection and sensing applications such as monitoring blood pressure or accumulating temperature readings. 
\begin{figure*}[h]
\centering
\includegraphics[width=0.85\textwidth,height=4cm]{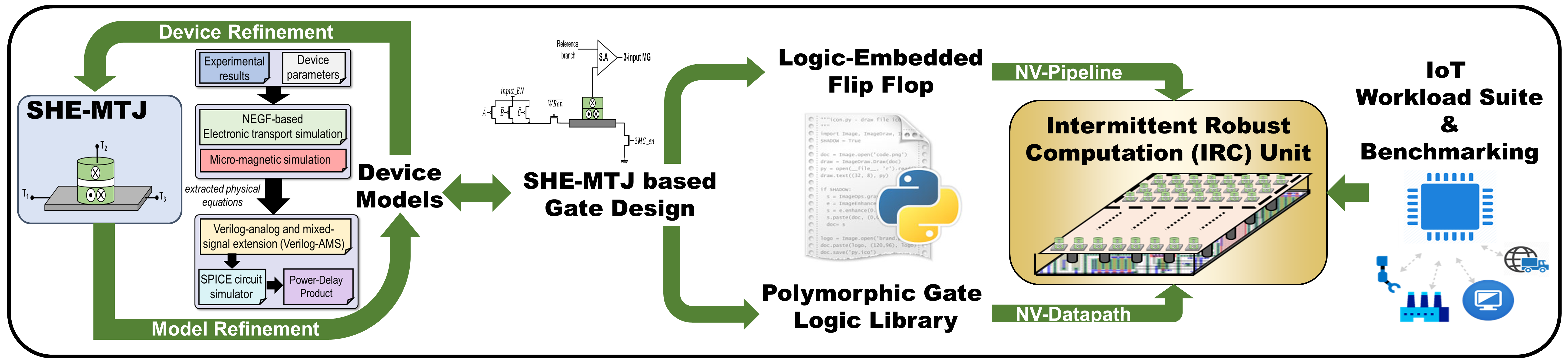}
\vspace{-1em}
\caption{Proposed cross-layer research on spin-based designs to attain energy harvesting-powered IoT.}
\label{fig1}
\vspace{-1.5em}
\end{figure*}

\subsection{Advancements Beyond Previous Works}
Table \ref{tb1} lists some of the prior efforts addressing the intermittency challenge facing energy-harvesting-powered designs. In \cite{mementos}, a traditional \emph{checkpointing} approach is utilized to ensure the accurate forward progress of computation, whereby any volatile execution context is proactively preserved in Non-Volatile Memory (NVM) prior to anticipated periods of power failure. A checkpointing approach may suffer from internal and external inconsistencies after each power loss. Internal inconsistency occurs when the execution context is partially-retained in NVM, while external inconsistency arises when the power failure occurs between two checkpoints. 

DINO \cite{DINO} innovated a checkpointing-based approach that utilizes non-volatile versioning to retain memory consistency, as delineated in Table \ref{tb1}. \emph{Duty Cycling with Scheduling} \cite{duty1} offers another approach for tolerating intermittence. In this method, critical states of the processor will be partially-retained before the power failure, then the device will enter an extremely-low power mode. However, this results in the full availability of the device only when power interruption is unlikely, which can incur relatively long sleeping periods due to the inevitable power outages in many energy harvesting-powered systems. Chain \cite{chain} is another model for programming intermittent devices, in which forward-progress is ensured at the task granularity level. It utilizes \emph{idempotent processing} concepts to make tasks restartable that never experience inconsistency to keep NVM consistent. In \cite{intermittency1}, a Non-Volatile MIPS Processor (NVP) is introduced in which specific blocks such as register files and pipeline registers were replaced by non-volatile elements. As listed in the last row of the Table \ref{tb1}, NVP utilizes a checkpointing approach to retain the processor volatile states resulting in possible above-mentioned internal and external inconsistencies in non-volatile elements. Advancing beyond previous intermittent processors which utilize NVM resources that are distinct from the processing datapath, we propose a new paradigm for energy-harvesting-powered processing, referred to as \emph{Intermittent Robust Computing} (IRC).

\section{Research Objectives}
In this project, an Intermittent Robust Computation (IRC) Unit cross-layer approach to energy-harvesting-powered processing, from device-level to architectural-level, is developed, as shown in Figure \ref{fig1}. It leverages non-volatility in spin-based devices when selectively-inserted into the datapath. Simulations and analyses attain the Research Objectives (ROs) below:

\emph{\underline{RO\#1}}: Construct open-source physics-based and compact models of novel spin-based devices designed for intermittent computation. Investigate spin-based device tradeoffs between write energy and volatility for various switching energy barriers. To explore the energy characteristics of spin-based VLSI circuits as well as innovate novel architectural schemes, Verilog-A, Matlab, and SPICE models will be developed enabling straightforward integration with VLSI circuits in SPICE-like platforms.

\emph{\underline{RO\#2}}: Design, simulate, and analyze Polymorphic Gate (PG) library and Logic-Embedded Flip-Flop (LE-FF) using the developed compact models to realize beyond-CMOS datapaths. Libraries containing a functionally-complete set of Boolean logic gates will be defined and populated using the developed device models. Moreover, the concept and design of the Logic-Embedded Flip-Flop (LE-FF) will be innovated. Extend the SHE-MTJ based design of the LE-FF from preliminary results with \textbf{15} ISCAS benchmark circuits simulated with a 40kT energy barrier, towards the LE-FF developed having significantly lower energy barriers. 

\emph{\underline{RO\#3}}: Develop Intermittent Robust Computation IP cores realizing normally-off computation via non-volatile datapaths using developed PG library and LE-FF. An NV-clustering methodology will be developed for targeted insertion of LE-FFs as new compact means to increase the functionality of pipeline registers. New algorithms will be developed to selectively utilize low energy barrier NV-PGs within the datapath to realize intermittent computation at reduced energy, while maintaining middleware coherence. To verify the functionality and demonstrate energy and performance characteristics, we implement ISCAS-89 benchmark circuits in commercial synthesis tools.

\section{Cross-Layer Co-Design Implementation of IRC}
\subsection{Device-level Approach to Intermittent Computation}
\begin{figure}[t]
\includegraphics[width=0.49\textwidth,height=6.5cm]{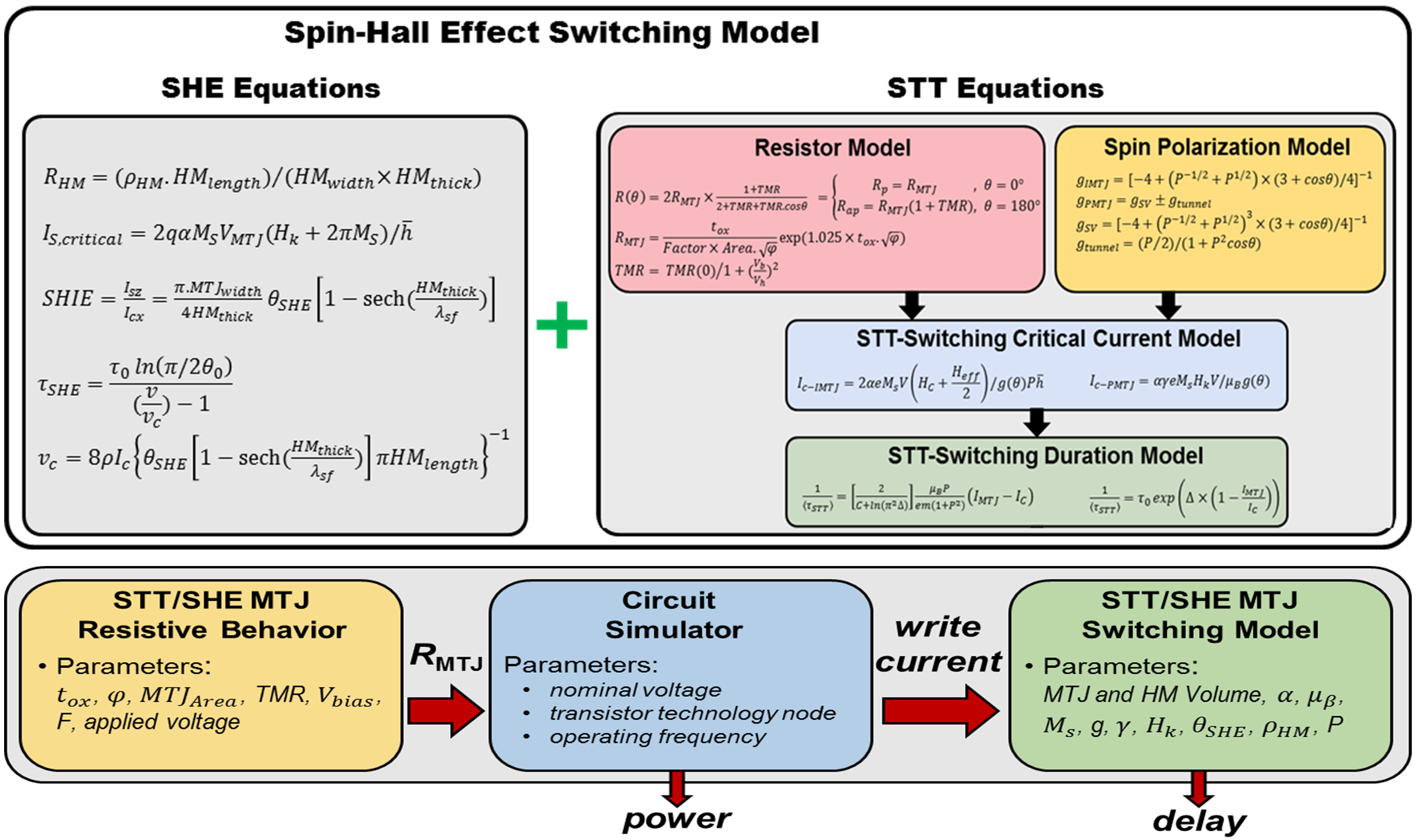}
\caption{Modeling and simulation process of STT/SHE MTJ devices.}
\label{fig2}
\vspace{-1em}
\end{figure}

\begin{figure}[b]
\includegraphics[width=0.49\textwidth]{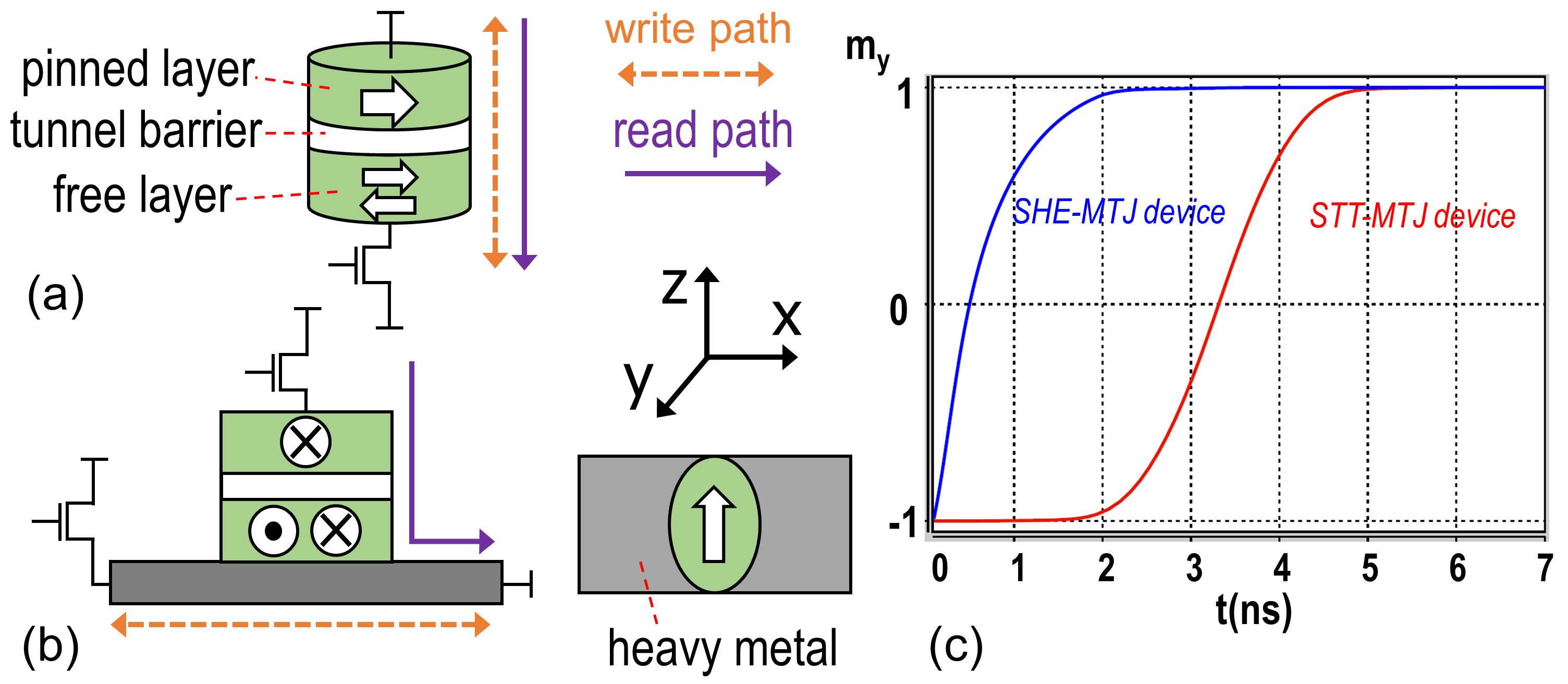}
\caption{(a) two-terminal In-plane MTJ structure, (b) three-terminal SHE-MTJ, vertical and top view (left and right, respectively), and (c) magnetization switching time for SHE- and STT- MTJs.}
\label{figsw}
\end{figure}

As shown in Figure \ref{fig2}, by integrating LLG Solver, Verilog-A, and SPICE models, a compact model for STT-IMTJ, STT-PMTJ \cite{armanDW}, and SHE-MTJ \cite{armanSHE} are developed to explore the energy and delay characteristics of spin-based VLSI circuits and to innovate novel architectural schemes utilizing non-volatile logic. They express the underlying both static and dynamic switching behavior and encapsulate their characteristics, while allowing straight-forward integration with VLSI circuits in SPICE-like platforms. To validate the model, several STT \& SHE -based memory elements \cite{SHE-LUT1} and functional building blocks \cite{armanDW} are designed and their functionalities are verified. Figure \ref{figsw} depicts structures of STT and SHE -MTJ along with  their switching time.

Due to the computational mechanism for spin-based devices, which is an accumulation-mode, spin-based devices can naturally function as a polymorphic gate (PG) and majority gate (MG). PGs/MGs can be cascaded to realize conjunctive/disjunctive Boolean gate realizations \cite{GLSVLSI}. For instance, by affixing one of the three inputs in ON or OFF states, then a 2-input OR gate or a 2-input AND gate can be realized, respectively. Therefore, we designed, simulated, and analyzed PG/MG libraries using the developed compact models. Libraries containing a functionally-complete set of Boolean logic gates are defined and populated using the developed device models.

\subsection{Circuit-level Approach to Intermittent Computation}
To achieve \emph{\underline{RO\#2}} spans the design and evaluation of Boolean logic gates realized by the modeled non-volatile spintronic devices. To implement a 3-input Non-Volatile Polymorphic Gates (NV-PGs), which is designed using SHE-MTJ devices, a pre-charge sense amplifier (PCSA) \cite{SA1} is utilized to sense the state of the SHE-MTJs. Reference MTJ dimensions are designed such that its resistance value in parallel configuration is between low resistance, R\textsubscript{Low}, and high resistance, R\textsubscript{High}, of the PG cells as elaborated by following equation, $R_{(P-REF)}\cong(R_{Low}+R_{High})/2$, where $R_{Low}=(R_{P-PG}+R_{HM})/2$ and $R_{High}=(R_{AP-PG}+R_{HM})/2$. The minimum current required for switching the state of the SHE-MTJ devices is called the critical current (IC), which is relative to the dimensions of the device. In an n-input NV-PG, the device is designed such that at least $(n-1)/2$ of the input transistors should be ON to produce a switching current amplitude greater than the critical current. We propose to research approaches by which NV-PGs can be cascaded to realize conjunctive/disjunctive Boolean gate realizations. For instance, by affixing one (or two) of the three (or five) input transistors in ON or OFF states upon demand during the circuit operation, then a 2(or 3)-input OR gate or a 2(or 3)-input AND gate can be realized, respectively.
\begin{figure}
\includegraphics[width=0.49\textwidth,height=7cm]{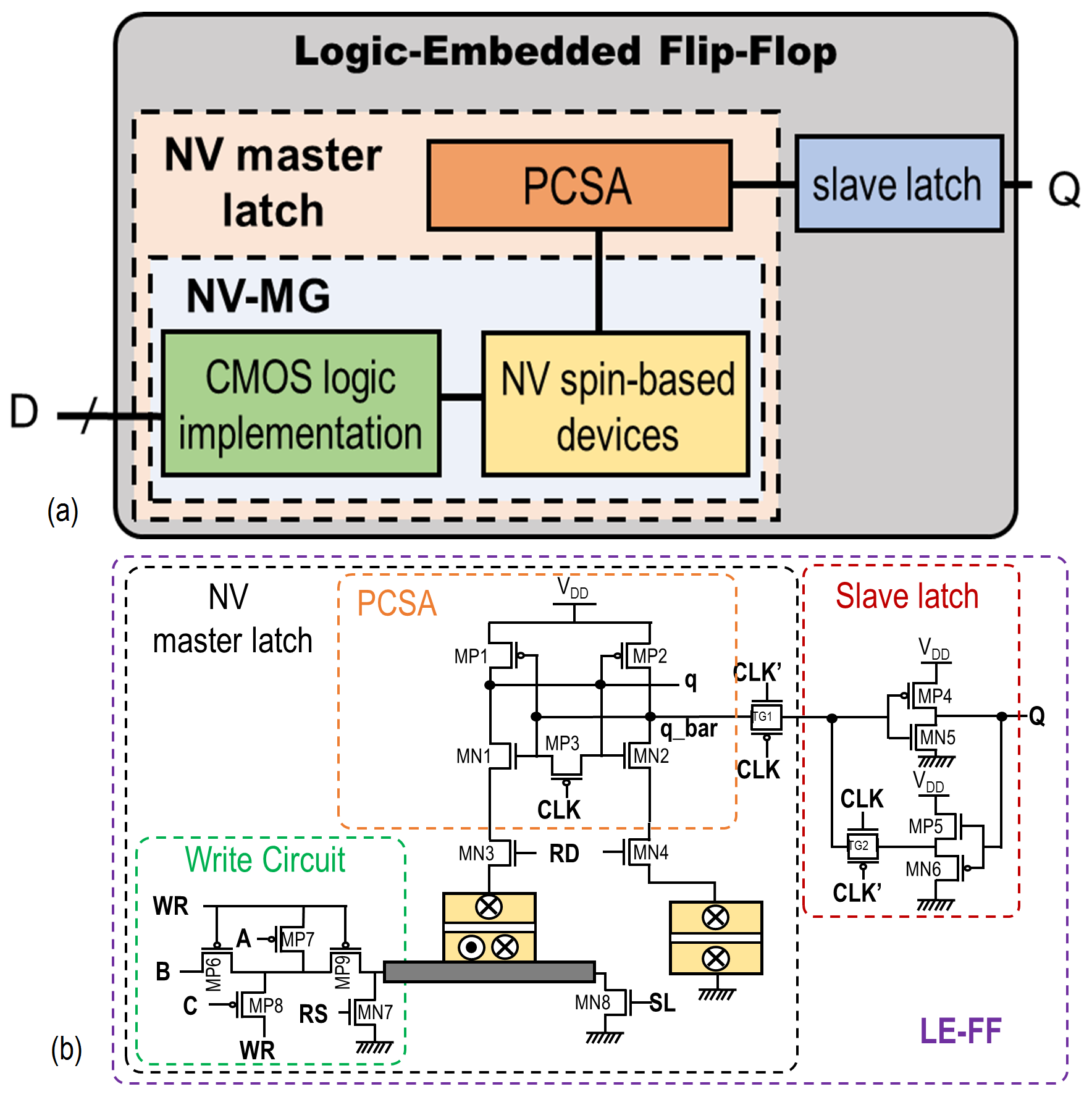}
\caption{(a)Schematic of proposed MG-based LE-FF, and, (b) circuit level design of proposed 3-input SHE-based LE-FF}
\label{fig3}
\vspace{-2em}
\end{figure}
\begin{figure*}
\centering
\includegraphics[width=0.83\textwidth,height=4.15cm]{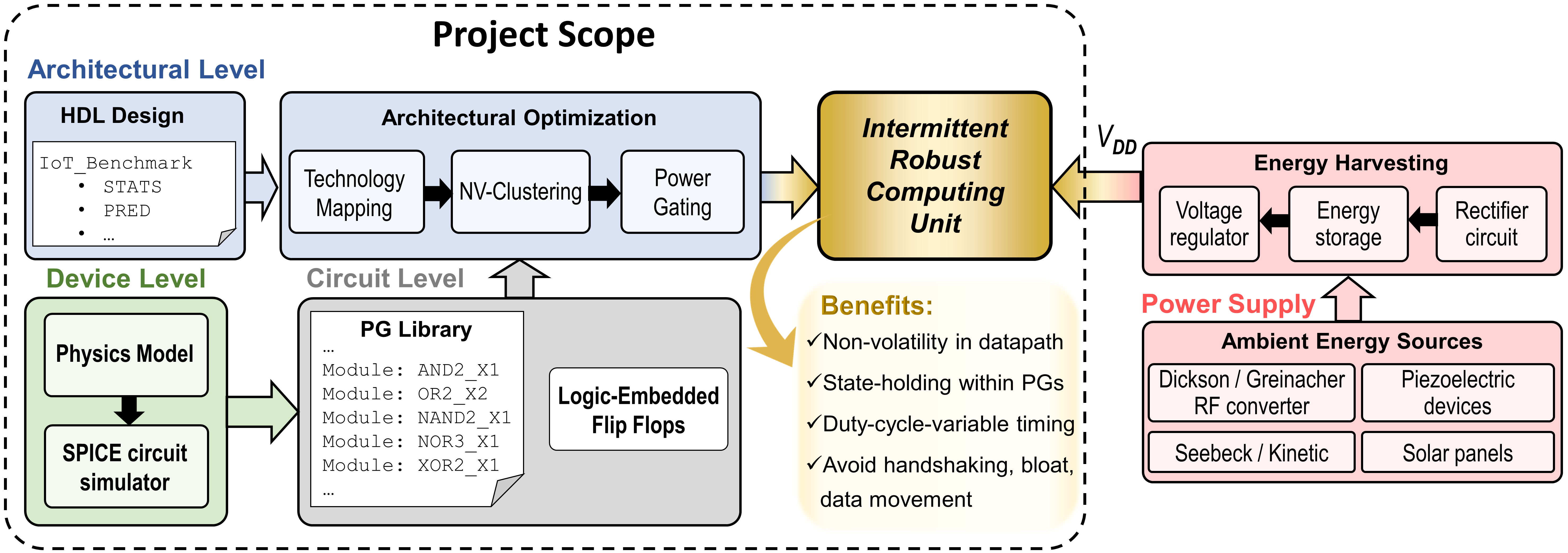}
\vspace{-0.5em}
\caption{Implementation methodology for RIC Unit powered by energy harvesting system.}
\label{fig5}
\vspace{-1em}
\end{figure*}
Next, a Logic-Embedded Flip-Flop (LE-FF) circuit is developed \cite{Roohi2018TC}. It is composed of an NV-PG based master latch using SHE-MTJ devices with 30kT-35kT energy barriers achieving retention times ranging from hours to days, as well as a CMOS-based slave latch, as shown in Figure \ref{fig3}a. An LE-FF has three different functional modes: (1) store mode, in which a targeted Boolean logic is implemented and stored in NV-PG; (2) standby mode, in which the power is OFF and data is held in the master latch due to the non-volatility feature of SHE-MTJs; and (3) sense mode, where the stored data in NV-PG is read and moved to the slave latch when the power is ON again. Our proposed LE-FF has two interesting features in comparison to the previously presented NV-FF designs: (a) in addition to storing a value with near-zero standby power, similar to the other NV-FFs, the LE-FF design is capable of intrinsically computing the primary Boolean expressions, resulting in area, complexity, and power reduction, (b) utilization of LE-FFs in large scale designs reduces their sensitive time (t\textsubscript{S}) to power failures. Sensitive time is defined as the duration of signal propagation between two NV elements including: (1) input registers and an NV-FF, (2) two NV-FFs, or (3) an NV-FF and output registers, in which if a power failure occurred, data will be lost, and rebooting and pipeline flushing is required. The vulnerability interval is expressed by t\textsubscript{S} = t\textsubscript{WR} + t\textsubscript{RD} + t\textsubscript{C}, where, t\textsubscript{WR} is the write operation time for the NV element, t\textsubscript{RD} is the switching time of CMOS-based slave latch, and t\textsubscript{C} is the delay of combinational circuits, the summation of all obtained sensitive times is defined as a \emph{Design Vulnerability Time} (DVT), according to which smaller DVT indicates higher tolerance to the power failure. The DVT of an integrated circuit can be reduced by replacing cones of gates and NV-FFs by LE-FFs, which increases the failure robustness. In order to design optimized NV architectures using the proposed LE-FF, we develop a systematic methodology, which incorporates all LE-FF features to design power-failure tolerant architectures. The proposed approach leverages the maximum capability of LE-FFs in terms of replacement and implementation steps.  

\subsection{Architecture-level Approach: IRC Unit }
The power profile of ambient energy sources imposes fundamental constraints on processing stability and duration. To achieve \emph{\underline{RO\#3}}, circuit-level results are extended towards benchmark studies corresponding to lifetime energy reduction and intermittent operational behavior demanded by IoT applications. Both goals can be achieved by selectively non-volatile datapaths using low energy barrier spin-based NV devices within the combinational circuits, as well as targeted insertion of LE-FFs as pipeline registers. Thus, the intermittent operation is supported without the burden of additional circuitry otherwise required for checkpoint-restore, backup, etc. Figure \ref{fig5} shows the VLSI synthesis methodology proposed herein for implementing logic circuits. The circuit-level simulation framework is utilized to develop a standard cell PG library containing a sufficient range of energy-efficient logic gates, as well as LE-FF circuits. We develop an NV-clustering methodology, which takes a Hardware Description Language (HDL) representation of a datapath and PG-based gate modules as its inputs and produces an optimized NV-enhanced datapath. A preliminary version of this algorithm is developed in Python, which explores the HDL of the logic circuit and finds the gates and pipeline registers that can be combined and replaced by spin-based LE-FF circuits, and the remainder of pipeline registers will be replaced by NV-FFs. In particular, 
the algorithm first finds all of the FFs in the design, and then checks the cone of logic gates connected to the inputs of the FFs. If each cone of gates meets the circuit-level criteria mentioned below, then the cone and its corresponding FF can be replaced by an LE-FF circuit. The three primary criteria are: (1) it should be possible to implement the cone of gates with a single PG, (2) fan-out of every gate in the cone should not exceed one, (3) none of the gates in the cone should be connected to the output of another FF \cite{Roohi2018TC}. 

\begin{figure*}
\includegraphics[width=0.98\textwidth,height=4.85cm]{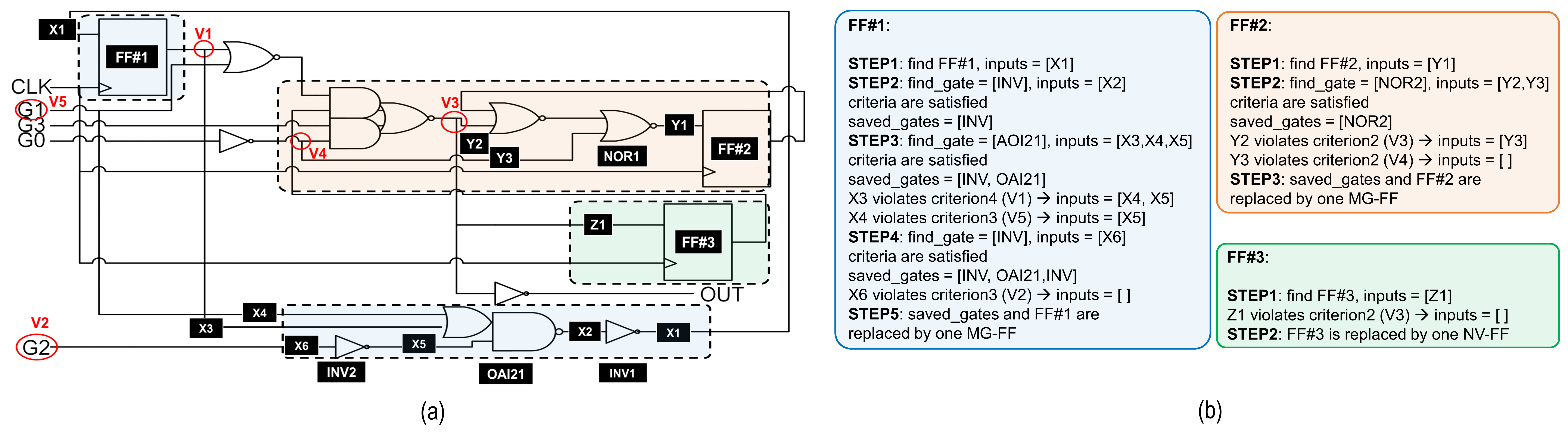}
\vspace{-1em}
\caption{(a) s27 schematic with highlighted FFs, and (b) all three applied NV-Clustering scheme.}
\label{fig6}\vspace{-1em}
\end{figure*}
\begin{figure}[t]
\includegraphics[width=0.48\textwidth,height=6.5cm]{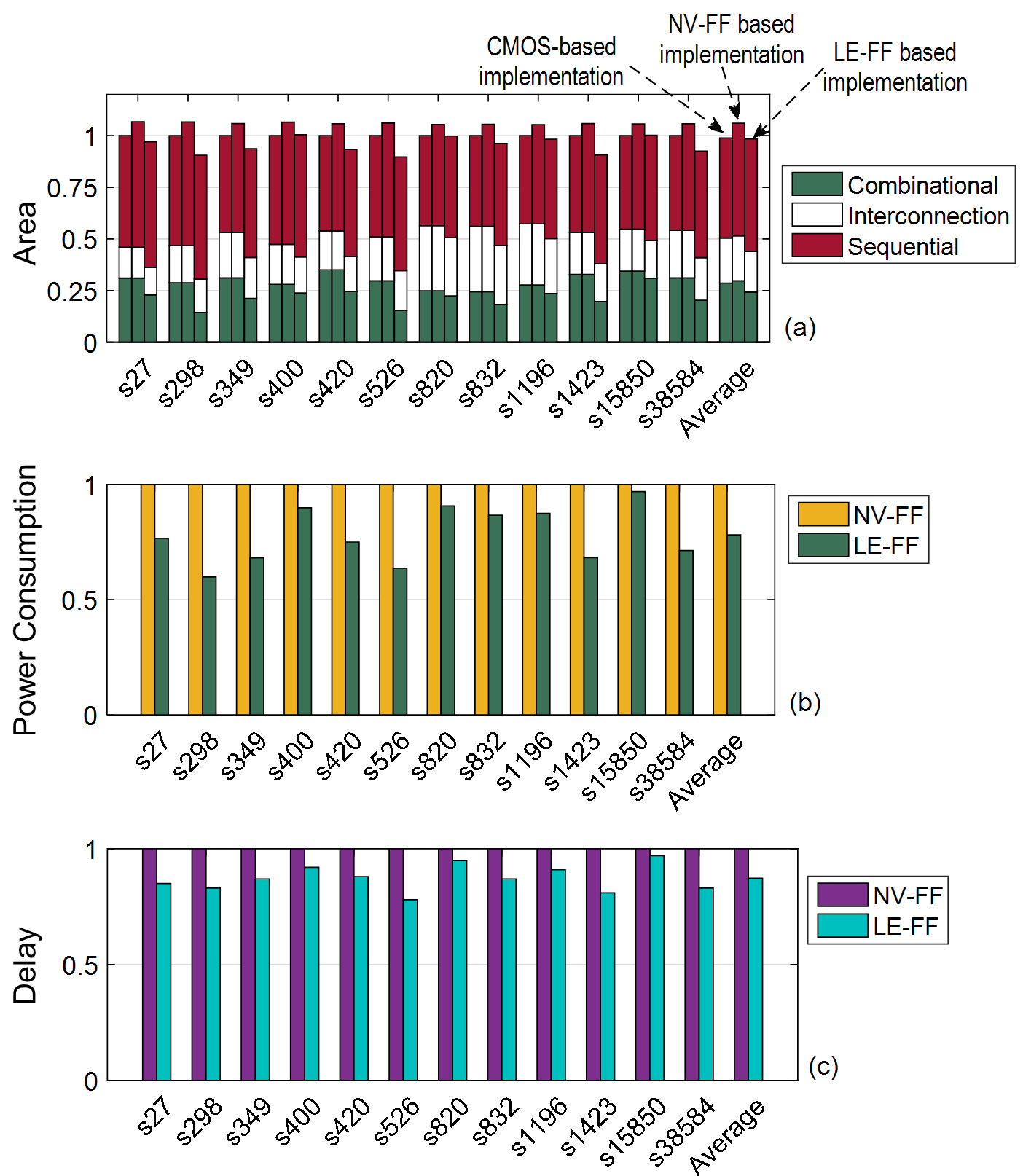}\vspace{-1em}
\caption{(a) s27 schematic with highlighted FFs, and (b) all three applied NV-Clustering scheme.}
\label{result}
\end{figure}

To exemplify the functionality of the proposed methodology, the s27 circuit from the ISCAS-89 benchmark is analyzed, as shown in Figure \ref{fig6}a. To estimate area, power, and delay improvements of our proposed approach, we have utilized a commercial synthesis tool, i.e. Synopsys Design Compiler, to map the input HDL to the developed spin-based PG library. The estimated results for 12 ISCAS-89 benchmark circuits are shown in Figure \ref{result}. Our proposed approach has achieved an average of 15\%, 22\%, 14\%, and 33\% improvements in terms of area, power, delay, and energy consumption, respectively, compared to the conventional intermittent computing circuits where all of the pipeline registers are replaced with NV-FFs. These estimated results are obtained by using SHE-MTJs with 40 kT energy barrier ($\Delta$), which can provide non-volatility for years. However, in the energy-harvesting-powered IoT devices, retention time in order of days and hours could be sufficient to achieve proper functionality. Therefore, the energy barrier of SHE-MTJ devices can be reduced to 30kT, realizing 25\% reduction in switching critical current ($I_C \propto \Delta$) and approximately 44\% decrease in energy consumption ($E \propto P \propto I^2$), while providing non-volatility for a few hours. Combining the above energy reductions to achieve at least 60\% improvements in energy consumption while reducing circuit lifetime energy-delay-product by at least 70\% compared to existing designs of comparable area motivates the proposed effort.

\section{Conclusion and Future Work}
In this project, the primary NV-clustering methodology was extended to realize the targeted insertion of 40-35kT SHE-MTJ devices within the combinational logic, enabling gate-level pipelining that is essential for energy harvesting applications. Next steps as future work, to avoid energy overhead caused by power gating control signals, we will create multiple gating domains including clustered gates, which can be simultaneously power-gated during circuit operation. Moreover, the state-holding characteristic of the spin-based devices facilitates ultra-fine-grained pipelining of logic paths without incurring additional overheads, which previously existed in conventional CMOS-based designs due to the presence of pipeline registers. Thus, we will research the benefit to operate intermittently without backing-up the processor state to separate non-volatile storage. Optimizations will be applied to improve energy and performance profiles of utilizing micro-benchmarks from a transportable suite of IoT benchmarks which will be developed. Finally, we will determine the minimum amount of the NV-PGs, which are required to be used in combinational circuits to ensure glitch-free PG-based IP Cores supporting intermittent computation with reduced energy consumption. The research insights gained regarding PG insertion will be generalized to arbitrary combinational logic expressions such as multi-modal Finite Impulse Response (FIR) filters with a backup power saving mode, which are representative case studies for IoT applications powered by energy harvesting applications.

%\section*{References}
{\small
\bibliographystyle{IEEEtran}  
\bibliography{IEEEabrv,./Reference}	

\begin{thebibliography}{10}
\providecommand{\url}[1]{#1}
\csname url@samestyle\endcsname
\providecommand{\newblock}{\relax}
\providecommand{\bibinfo}[2]{#2}
\providecommand{\BIBentrySTDinterwordspacing}{\spaceskip=0pt\relax}
\providecommand{\BIBentryALTinterwordstretchfactor}{4}
\providecommand{\BIBentryALTinterwordspacing}{\spaceskip=\fontdimen2\font plus
\BIBentryALTinterwordstretchfactor\fontdimen3\font minus
  \fontdimen4\font\relax}
\providecommand{\BIBforeignlanguage}[2]{{%
\expandafter\ifx\csname l@#1\endcsname\relax
\typeout{** WARNING: IEEEtran.bst: No hyphenation pattern has been}%
\typeout{** loaded for the language `#1'. Using the pattern for}%
\typeout{** the default language instead.}%
\else
\language=\csname l@#1\endcsname
\fi
#2}}
\providecommand{\BIBdecl}{\relax}
\BIBdecl

\bibitem{moore}
R.~R. Schaller, ``Moore's law: past, present and future,'' \emph{IEEE
  spectrum}, vol.~34, no.~6, pp. 52--59, 1997.

\bibitem{ITRS}
``The international technology roadmap for semiconductors,'' \emph{Available
  at: (http://www.itrs.net)}, 2015.

\bibitem{armanDW}
A.~Roohi, R.~Zand, and R.~F. DeMara, ``A tunable majority gate-based full adder
  using current-induced domain wall nanomagnets,'' \emph{IEEE Transactions on
  Magnetics}, vol.~52, no.~8, pp. 1--7, 2016.

\bibitem{armanSHE}
A.~Roohi, R.~Zand, D.~Fan, and R.~F. DeMara, ``Voltage-based concatenatable
  full adder using spin hall effect switching,'' \emph{IEEE Transactions on
  Computer-Aided Design of Integrated Circuits and Systems}, vol.~36, no.~12,
  pp. 2134--2138, 2017.

\bibitem{SHE-adv2}
S.~Manipatruni, D.~E. Nikonov, and I.~A. Young, ``Energy-delay performance of
  giant spin hall effect switching for dense magnetic memory,'' \emph{Applied
  Physics Express}, vol.~7, no.~10, p. 103001, 2014.

\bibitem{SHE-adv3}
D.~Fan, S.~Maji, K.~Yogendra, M.~Sharad, and K.~Roy, ``Injection-locked spin
  hall-induced coupled-oscillators for energy efficient associative
  computing,'' \emph{IEEE Transactions on Nanotechnology}, vol.~14, no.~6, pp.
  1083--1093, 2015.

\bibitem{mementos}
B.~Ransford, J.~Sorber, and K.~Fu, ``Mementos: System support for long-running
  computation on rfid-scale devices,'' in \emph{ACM SIGARCH Computer
  Architecture News}, vol.~39, no.~1.\hskip 1em plus 0.5em minus 0.4em\relax
  ACM, 2011, pp. 159--170.

\bibitem{duty1}
D.~Balsamo, A.~S. Weddell, G.~V. Merrett, B.~M. Al-Hashimi, D.~Brunelli, and
  L.~Benini, ``Hibernus: Sustaining computation during intermittent supply for
  energy-harvesting systems,'' \emph{IEEE Embedded Systems Letters}, vol.~7,
  no.~1, pp. 15--18, 2015.

\bibitem{DINO}
B.~Lucia and B.~Ransford, ``A simpler, safer programming and execution model
  for intermittent systems,'' \emph{ACM SIGPLAN Notices}, vol.~50, no.~6, pp.
  575--585, 2015.

\bibitem{chain}
A.~Colin and B.~Lucia, ``Chain: tasks and channels for reliable intermittent
  programs,'' \emph{ACM SIGPLAN Notices}, vol.~51, no.~10, pp. 514--530, 2016.

\bibitem{intermittency1}
K.~Ma, Y.~Zheng, S.~Li, K.~Swaminathan, X.~Li, Y.~Liu, J.~Sampson, Y.~Xie, and
  V.~Narayanan, ``Architecture exploration for ambient energy harvesting
  nonvolatile processors,'' in \emph{High Performance Computer Architecture
  (HPCA), 2015 IEEE 21st International Symposium on}.\hskip 1em plus 0.5em
  minus 0.4em\relax IEEE, 2015, pp. 526--537.

\bibitem{s-all}
S.~Bandyopadhyay and A.~P. Chandrakasan, ``Platform architecture for solar,
  thermal, and vibration energy combining with mppt and single inductor,''
  \emph{IEEE Journal of Solid-State Circuits}, vol.~47, no.~9, pp. 2199--2215,
  2012.

\bibitem{new-computing}
B.~P. Rao, P.~Saluia, N.~Sharma, A.~Mittal, and S.~V. Sharma, ``Cloud computing
  for internet of things \& sensing based applications,'' in \emph{Sensing
  Technology (ICST), 2012 Sixth International Conference on}.\hskip 1em plus
  0.5em minus 0.4em\relax IEEE, 2012, pp. 374--380.

\bibitem{medical1}
M.~Imran, ``Energy harvesting mechanism for medical devices,'' May~5 2015, uS
  Patent 9,026,212.

\bibitem{aerospace}
S.~P. Beeby, M.~J. Tudor, and N.~White, ``Energy harvesting vibration sources
  for microsystems applications,'' \emph{Measurement science and technology},
  vol.~17, no.~12, p. R175, 2006.

\bibitem{IoT}
M.~Gorlatova, J.~Sarik, G.~Grebla, M.~Cong, I.~Kymissis, and G.~Zussman,
  ``Movers and shakers: Kinetic energy harvesting for the internet of things,''
  in \emph{ACM SIGMETRICS Performance Evaluation Review}, vol.~42, no.~1.\hskip
  1em plus 0.5em minus 0.4em\relax ACM, 2014, pp. 407--419.

\bibitem{medical2}
A.~Poor, ``Reaping the energy harvest [resources],'' \emph{IEEE Spectrum},
  vol.~52, no.~4, pp. 23--24, 2015.

\bibitem{SHE-LUT1}
R.~Zand, A.~Roohi, D.~Fan, and R.~F. DeMara, ``Energy-efficient nonvolatile
  reconfigurable logic using spin hall effect-based lookup tables,'' \emph{IEEE
  Transactions on Nanotechnology}, vol.~16, no.~1, pp. 32--43, 2017.

\bibitem{GLSVLSI}
A.~{Roohi}, R.~{Zand}, and R.~F. {DeMara}, ``Logic-encrypted synthesis for
  energy-harvesting-powered spintronic-embedded datapath design,'' ser. GLSVLSI
  '18.\hskip 1em plus 0.5em minus 0.4em\relax ACM, 2018, pp. 9--14.

\bibitem{SA1}
W.~Zhao, C.~Chappert, V.~Javerliac, and J.-P. Noziere, ``High speed, high
  stability and low power sensing amplifier for mtj/cmos hybrid logic
  circuits,'' \emph{IEEE Transactions on Magnetics}, vol.~45, no.~10, pp.
  3784--3787, 2009.

\bibitem{Roohi2018TC}
A.~Roohi and R.~F. DeMara, ``Nv-clustering: Normally-off computing using
  non-volatile datapaths,'' \emph{IEEE Transactions on Computers}, vol.~67,
  no.~7, pp. 949--959, July 2018.

\end{thebibliography}
}

\end{document}